%% file: main.tex
\documentclass[sigconf]{acmart}
\AtBeginDocument{%
  }

\newcommand{\framework}{\textsc{RelianceScope}}

\usepackage{multirow}
\usepackage{twemojis}
\usepackage{enumitem}
\usepackage{listings}
\usepackage{url}
\usepackage{tabularx}
\usepackage{balance}

\copyrightyear{2026}
\acmYear{2026}
\setcopyright{cc}
\setcctype{by}
\acmConference[L@S '26]{Proceedings of the Thirteenth ACM Conference on Learning @ Scale}{June 29-July 03, 2026}{Seoul, Republic of Korea}
\acmBooktitle{Proceedings of the Thirteenth ACM Conference on Learning @ Scale (L@S '26), June 29-July 03, 2026, Seoul, Republic of Korea}
\acmDOI{10.1145/3774398.3811612}
\acmISBN{979-8-4007-2293-6/2026/06}

\begin{document}

\title[\framework{}: A Framework for Analyzing Students' Reliance on AI]{\framework{}: An Analytical Framework for Examining Students' Reliance on Generative AI Chatbots in Problem Solving}

\author{Hyoungwook Jin}
\email{jinhw@umich.edu}
\orcid{0000-0003-0253-560X}
\affiliation{%
  \institution{University of Michigan}
  \city{Ann Arbor}
  \state{Michigan}
  \country{USA}
}

\author{Minju Yoo}
\email{minjuyoo@kaist.ac.kr}
\orcid{0009-0000-2964-0464}
\affiliation{%
  \institution{KAIST}
  \city{Daejeon}
  \country{Republic of Korea}
}

\author{Jieun Han}
\email{jieun_han@kaist.ac.kr}
\orcid{0009-0003-7740-517X}
\affiliation{%
  \institution{KAIST}
  \city{Daejeon}
  \country{Republic of Korea}
}

\author{Zixin Chen}
\email{zchendf@connect.ust.hk}
\orcid{0000-0001-8507-4399}
\affiliation{%
  \institution{HKUST}
  \city{Hongkong}
  \country{China}
}

\author{So-Yeon Ahn}
\email{ahnsoyeon@kaist.ac.kr}
\orcid{0000-0003-2718-0999}
\affiliation{%
  \institution{KAIST}
  \city{Daejeon}
  \country{Republic of Korea}
}

\author{Xu Wang}
\email{xwanghci@umich.edu}
\orcid{0000-0001-5551-0815}
\affiliation{%
  \institution{University of Michigan}
  \city{Ann Arbor}
  \state{Michigan}
  \country{USA}
}

\renewcommand{\shortauthors}{Hyoungwook Jin et al.}

\begin{abstract}
\input{section/00_abstract}
\end{abstract}


\begin{teaserfigure}
  \includegraphics[width=\textwidth]{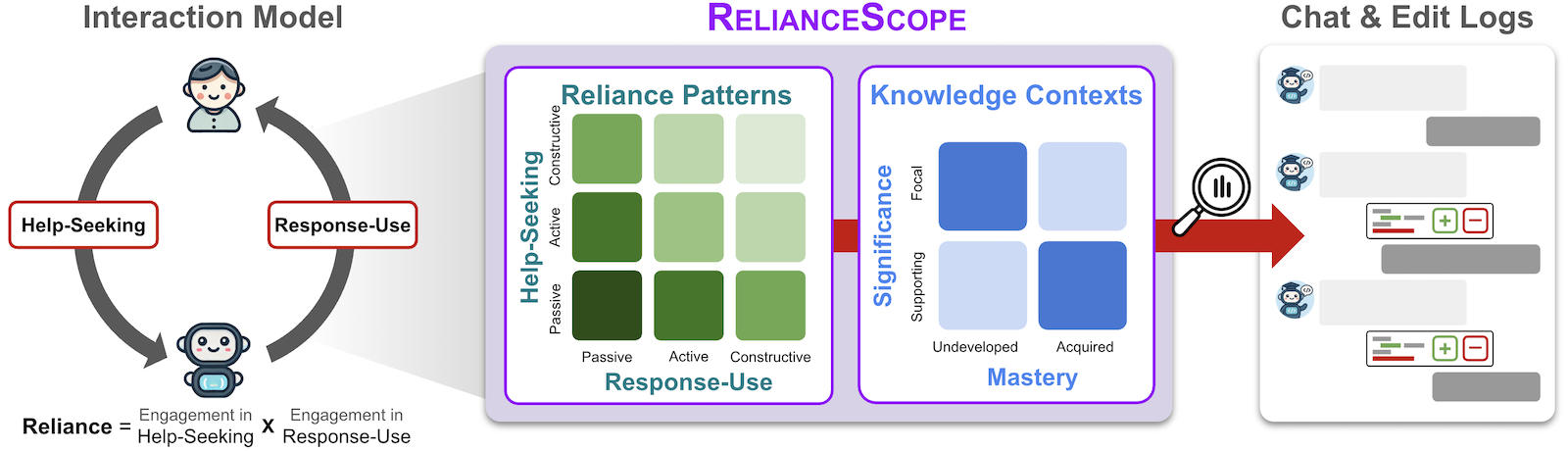}
  \caption{Combined engagement in help-seeking and response-use is critical for examining students' reliance on AI during chatbot-assisted problem-solving. \framework{} is a framework for analyzing these reliance patterns from chat and artifact edit logs through the lens of knowledge contexts.}
  \Description{A diagram illustrating how the Interaction Model and RelianceScope framework are used to analyze chat and artifact edit logs. On the left, the Interaction Model depicts a cyclical relationship between the student and the chatbot, labeled Help-Seeking and Response-Use, with Reliance defined as the product of engagement in both. In the center, RelianceScope provides two analytical lenses: Reliance Patterns, a three-by-three grid with Help-Seeking modes (Passive, Active, Constructive) on the vertical axis and Response-Use modes (Passive, Active, Constructive) on the horizontal axis; and Knowledge Contexts, a two-by-two grid with Significance (Focal, Supporting) on the vertical axis and Mastery (Undeveloped, Acquired) on the horizontal axis. On the right, these lenses are applied to Chat and Edit Logs.}
  \label{fig:teaser}
\end{teaserfigure}


\maketitle

\input{section/10_introduction}
\input{section/20_related_work}
\input{section/30_framework}
\input{section/40_study}
\input{section/50_results}
\input{section/60_discussion}
\input{section/70_limitation}

\begin{acks}
This work is funded by NSF Grant IIS-2442990. The findings and conclusions expressed in this material are those of the author(s) and do not necessarily reflect the views of the National Science Foundation.
\end{acks}

\bibliographystyle{ACM-Reference-Format}
\balance
\bibliography{base}



\end{document}

%% file: section/00_abstract.tex
Generative AI chatbots enable personalized problem-solving, but effective learning requires students to self-regulate both how they seek help and how they use AI-generated responses. Considering engagement modes across these two actions reveals nuanced reliance patterns: for example, a student may actively engage in help-seeking by clearly specifying areas of need, yet engage passively in response-use by copying AI outputs, or vice versa. However, existing research lacks systematic tools for jointly capturing engagement across help-seeking and response-use, limiting the analysis of such reliance behaviors. We introduce \framework{}, an analytical framework that characterizes students' reliance on chatbots during problem-solving. \framework{} (1) operationalizes reliance into nine patterns based on combinations of engagement modes in help-seeking and response-use, and (2) situates these patterns within a knowledge-context lens that accounts for students' prior knowledge and the instructional significance of knowledge components. Rather than prescribing optimal AI use, the framework enables fine-grained analysis of reliance in open-ended student–AI interactions. As an illustrative application, we applied \framework{} to analyze chat and code-edit logs from 79 college students in a web programming course. Results show that active help-seeking is associated with active response-use, whereas reliance patterns remain similar across knowledge mastery levels. Students often struggled to articulate their knowledge gaps and to adapt AI responses. Using our annotated dataset as a benchmark, we further demonstrate that large language models can reliably detect reliance during help-seeking and response-use. We conclude by discussing the implications of \framework{} and the design guidelines for AI-supported educational systems.

%% file: section/10_introduction.tex
\section{Introduction}
Students' reliance on generative AI technologies has become a central concern in education research and practice~\cite{bastani2024generative}. Unlike conventional computer-assisted tutoring systems, the generative-AI-based chatbots enable highly flexible and personalized interactions~\cite{adamopoulou2020overview}, allowing students to pose questions that extend beyond instructor-designed learning pathways~\cite{gomes2025ai, openai2025}. Yet this open-ended interaction places substantial demands on students' ability to use AI productively~\cite{aleven2000limitations, azevedo2004role, baker2008students, amoozadeh2024student, chen2024plagiarism, zimmerman2002becoming}. Students can easily over-rely on chatbots---for example, by requesting complete solutions and copying generated outputs with minimal learning. Such emerging usage patterns call for the systematic capture of when and how students rely on chatbots, as well as an understanding of what constitutes productive use of generative AI~\cite{kosmyna2025your, shen2026ai}.

Students' cognitive engagement is particularly important for characterizing their reliance on a chatbot, given their interactions with it. We conceptualize student–chatbot interaction during problem-solving as a loop inspired by Norman's execution-evaluation model \cite{norman2013design}: a student seeks help from a chatbot and then uses the chatbot's response to modify a learning artifact. Within this loop, students can rely on chatbots at two key points. First, during \textit{help-seeking}, students may rely on chatbots by delegating decisions about the type of help to receive (e.g., ``help me'' vs. ``give me an example instead of the answer'')~\cite{reeve2011agency, satyanarayan2024intelligence}. Second, during \textit{response-use}, students may rely on chatbots by delegating the completion of the learning task (e.g., copying a response vs. writing independently)~\cite{chi2014icap, weatherholtz2025cognitive}. Together, the engagement modes during help-seeking and response-use provide a holistic view of how students rely on chatbots.

However, existing research lacks a unified framework for examining reliance across both help-seeking and response-use within student-chatbot interactions. Prior work has primarily analyzed these two processes in isolation~\cite{demszky2021measuring, sheese2024patterns, han2024recipe4u, mcnichols2025studychat, weatherholtz2025cognitive}. While such focused analyses provide valuable insights, they overlook how contrasting reliance behaviors combined within an interaction loop shape learning. For instance, a student may actively engage in help-seeking by asking personalized questions beyond learning objectives, yet exhibit minimal engagement in response-use without further actions. Conversely, another student may passively follow a chatbot's guidance but constructively engage with it by adapting responses rather than copying them verbatim. Can an isolated examination distinguish the reliance patterns of these interactions? How would this nuanced, joint reliance correlate with learning?

We propose \framework{} (Figure~\ref{fig:teaser}), an analytical framework for capturing and examining students' joint reliance on chatbots during help-seeking and response-use. We defined three engagement modes~\cite{chi2014icap} for each, yielding 3 × 3 reliance patterns at individual student-chatbot interaction units. Besides, compared with prior approaches, \framework{} further introduces a 2 × 2 knowledge-context lens that accounts for students' mastery of the knowledge component and its instructional significance. The knowledge context enables more contextualized interpretations of reliance behaviors. Last but not least, analyses using \framework{} are both scalable and analytically rich, as the framework relies on automatically collectible chat logs and artifact-edit logs without disrupting students' learning processes.

As a proof of concept, we developed a data collection system and used \framework{} to analyze students' reliance on chatbots in a college-level web programming course. In the course, students learned Vue.js by building a To-Do application with a generative AI chatbot. We created a web-based learning system and a chatbot to collect chat logs, code-edit logs, copy logs, self-regulation measures, and pre- and post-test assessments from 79 students. The learning activities were structured into incremental steps, enabling students to engage with the chatbot for each pre-defined knowledge component. This design allowed us to capture reliance at fine-grained interaction units spanning multiple conversational turns. 

We manually annotated reliance patterns for the collected data and conducted quantitative and qualitative analyses. Results show that higher engagement in help-seeking correlates with higher engagement in response-use, and that students exhibit similar reliance patterns when working with acquired or undeveloped knowledge. Students were highly reliant on the chatbot, the most reliant pattern accounting for 44\%; survey responses indicate that this reliance was associated with difficulties decomposing help into specific knowledge gaps and integrating AI-generated code into their own code. Using the human-annotated dataset as a benchmark, we further demonstrate that large language models can achieve moderate performance in detecting reliance patterns defined by \framework{}, with particularly high accuracy for passive help-seeking ($F1 = .807$) and response-use ($F1 = .760$).

To the best of our knowledge, this work presents the first interaction-centric framework that integrates help-seeking and response-use to define reliance in generative AI chatbot-assisted learning. We envision that the fine-grained, context-sensitive, and non-intrusive design of \framework{} will inform future investigations of reliance and guide the design of educational systems that better support self-regulated learners.

This paper makes the following contributions:
\begin{itemize}
    \item \framework{}, an analytical framework for classifying and contextualizing students' reliance on AI using both chatbot–student chat logs and artifact-edit logs.

    \item A generative AI chatbot-assisted learning system deployed in a college-level web programming course that collects chat logs, code edit logs, and copy logs.

    \item A publicly available dataset\footnote{\url{https://osf.io/27ec5/overview?view_only=a8234a17f908464297d35d5ca1ef476c}} comprising 1,362 chat logs annotated with reliance patterns, 2,708 code edit logs, pre- and post-assessments, and self-reported self-regulation measures from 79 students that can be used for educational data mining and reliance pattern classifier training.

    \item A quantitative analysis using \framework{} to examine the distribution of students' reliance patterns on a programming learning activity across varying knowledge contexts.
    
    \item A qualitative analysis of survey responses using the interaction model as a lens to characterize students' challenges in managing reliance during help-seeking and response-use.
\end{itemize}

%% file: section/20_related_work.tex
\section{Related Work}
We outline the reliance on generative AI in relation to self-regulation and review existing tools for capturing reliance.

\subsection{Reliance in Generative AI-Assisted Learning}
Research on AI reliance in learning has largely been framed from two perspectives: critical thinking and self-regulation. From a critical-thinking perspective, prior work emphasizes learners' ability to evaluate the accuracy of AI-generated outputs and filter out misleading or incorrect information~\cite{buccinca2021trust, zhai2024effects, zheng2025students, li2025can}. On the other hand, research grounded in self-regulation, including our work, focuses on how learners set goals with motivation, monitor their progress metacognitively, and proactively personalize their use of AI~\cite{zimmerman2002becoming, pitts2025students}. From this perspective, over-reliance arises when learners lose agency over their learning and choose ineffective engagement with the target knowledge.

Generative AI–based learning environments further amplify the importance of self-regulation. Unlike traditional tutoring systems that operate within predefined student models and corresponding guidance~\cite{graesser2005autotutor, ritter2007cognitive}, chatbot-based systems provide open-ended, generative assistance across a wide range of tasks, affording learners substantial freedom---but also greater responsibility~\cite{prather2024widening}---in determining how and when to rely on AI. The open-ended interaction patterns with AI complicate the analysis and understanding of its impact on learning. Empirical findings in this space remain mixed: some studies report learning gains without clear evidence of deskilling~\cite{kazemitabaar2023studying, lira2025learning}, while others document risks such as metacognitive laziness~\cite{fan2025beware}, plagiarism~\cite{chen2024plagiarism}, and short-lived retention~\cite{bastani2024generative, zhou2025impact}. 

Together, these mixed findings may suggest that the impact of AI reliance depends on contextual factors and interaction patterns that existing measures do not fully capture, underscoring the need for standardized and context-sensitive analytical tools~\cite{ibrahim2025towards}. \framework{} addresses this gap by conceptualizing reliance along the dimensions of help-seeking and response-use---perspectives largely unexplored jointly in prior work---and by introducing a knowledge-context lens that enables systematic examination of when different reliance patterns emerge and affect learning.

\subsection{Capturing Reliance on AI}
Capturing how learners rely on AI is critical for understanding the role of chatbots. Prior work has proposed taxonomies to characterize reliance from different perspectives. Many studies categorize reliance by examining the content and intent of learners' messages~\cite{graesser1994question, yang2024debugging, sheese2024patterns, han2024recipe4u, lyu2024evaluating, mcnichols2025studychat}, help-seeking behavior~\cite{nelson1981help, karabenick2003seeking, kazemitabaar2023studying, myung2026scaffolding}, or patterns of social interaction~\cite{borchers2024combining}. Others move beyond individual message turns to identify broader reliance profiles~\cite{guner2025ai, stojanov2024university, hao2025student}. 

Effectively applying these taxonomies requires attention to context beyond chat classification. Prior work has incorporated artifact edit logs and even video recordings of learner behavior to observe moments of reliance that are not visible in chat alone~\cite{amoozadeh2024student, sun2024would}. Others have analyzed tutor-learner dialogues from multiple dimensions, including modeling the dynamic flow of self-regulation over multiple turns~\cite{naim2025towards, yang2025beyond, he2025towards}. In addition to behavioral traces, some studies use self-reporting methods to complement performance measures with learners' subjective awareness of reliance~\cite{hou2025measuring, shoufan2023exploring}. More intrusive qualitative methods, such as think-aloud protocols, have also been used to trace the unfolding phases of self-regulation during AI-assisted learning~\cite{zhang2024using}. 

Prior work faces a fundamental trade-off between scalability and analytical depth: large-scale studies rely on chat logs alone, limiting insight into students' response-use, whereas deeper analyses incorporate richer contextual data but are constrained by small sample sizes. To capture reliance holistically across help-seeking and response-use, a method must balance scalability and contextual richness. Our data collection system and \framework{} offer a practical middle ground by leveraging students' artifact edit logs for contextual, yet automatable analysis.

%% file: section/30_framework.tex
\section{\framework{}}
\framework{} is a novel analytical framework for fine-grained investigation of students' reliance on free-form, conversational AI in problem-solving activities. Learning science and engineering researchers can use the framework to investigate students' patterns, points, and challenges of reliance in their target learning activities.

We designed \framework{} specifically for conversational AI tutors, with which students have different interactions from tutors in intelligent tutoring systems and classrooms. With conversational AI, students can ask questions in free-form natural language and use their tutor's responses freely, making content and contextual analysis complicated but crucial. To this end, we created the framework with three design goals in mind.

\textbf{Fine-grained.} We view students' reliance on AI as a spectrum rather than a binary classification. While the extremes of reliance are obvious, the key to understanding its impact on learning lies in identifying the middle ground~\cite{he2025fine}, where students' reliance during help-seeking and response-use diverge. To distinguish nuances in students' reliance, we propose bidimensional patterns of reliance based on their interactions with AI.

\textbf{Context-sensitive.} We are particularly interested in capturing how students' reliance on generative AI varies across learning contexts. For instance, learners may rely more on AI when limited prior knowledge constrains their ability to actively engage in an activity. Understanding how knowledge mastery and instructional significance shape reliance patterns can inform the design of adaptive generative-AI tutoring systems. Our framework leverages these knowledge contexts as a lens to enhance its analytical capabilities.

\textbf{Nonintrusive.} Our framework captures students' natural interactions with AI without introducing disruptive data-collection interventions, such as intermittent surveys or think-aloud protocols~\cite{han2024recipe4u, zhang2024using}. Instead, it leverages chat logs and artifact edit logs to obtain rich analytical signals while preserving authentic learning behavior. This nonintrusive design enhances compatibility across diverse learning activities. We provide our data-collection system as an illustrative implementation.

\input{table/codebook_help_seeking}

\input{table/codebook_response_use}

\subsection{Components of the Framework}
\framework{} captures reliance patterns at the level of interaction segments (see Figure~\ref{fig:example_conversation}). An \textbf{interaction segment} comprises a sequence of contiguous messages, along with the corresponding record of a student's work on the learning activity (e.g., code edit history), all associated with the same knowledge component (e.g., concept of for loop, Python loop syntax in an introductory programming class). An interaction segment may be brief, consisting of a single student request and the chatbot's response, or it may span multiple turns with follow-up questions. Researchers and educators can identify interaction segments by associating each message with a knowledge component and grouping contiguous messages that address the same knowledge component. We adopt interaction segments as the unit of analysis because they represent the atomic unit of student-chatbot interaction around a specific knowledge component and can capture reliance patterns across multi-turn interactions.

\framework{} has two components for classifying each interaction segment---reliance patterns and knowledge contexts (Figure~\ref{fig:teaser}). \textbf{Reliance patterns} include 9 categories that capture students' varying engagement modes with AI. \textbf{Knowledge contexts} include 4 categories that describe students' knowledge mastery and the relevance of each knowledge component to core learning objectives. Together, reliance patterns serve as the primary lens for quantifying reliance, while knowledge contexts provide an additional layer that explains when particular reliance patterns emerge.

\subsubsection{Reliance Patterns}
We adopted a human-computer interaction perspective to frame reliance occurring in interactions between students and chatbots. We conceptualize an interaction model between a student and a chatbot to define the reliance patterns (Figure~\ref{fig:teaser}), adapted from Norman's execution–evaluation model~\cite{norman2013design}. Our model characterizes the interaction as a two-stage cycle: when students encounter difficulties during a learning activity (e.g., a programming assignment), they ask a chatbot for help and subsequently integrate its responses into their learning artifacts (e.g., code). In the model, \textbf{help-seeking} corresponds to Norman's \textit{bridge of execution}, through which students articulate goals and specify actions to the chatbot. \textbf{Response-use} corresponds to the \textit{bridge of evaluation}, in which students interpret the chatbot's output and decide whether additional interaction is needed or whether their goal has been achieved.

Within each interaction cycle, students make self-regulatory decisions about how deeply to engage in both actions: (1) in help-seeking, whether to precisely specify the content and form of assistance or defer those decisions to the AI; and (2) in response-use, whether to directly copy AI-generated outputs or actively adapt them into their own work. We categorize engagement modes in each process into three modes---passive, active, and constructive---informed by Chi's ICAP framework of cognitive engagement~\cite{chi2014icap} and prior work on student agency~\cite{reeve2011agency}. 

The primary distinction between passive and active modes lies in the level of cognitive effort required to formulate requests and integrate responses (see Tables~\ref{table:codebook_help_seeking} and~\ref{table:codebook_response_use}). The \textbf{\textit{passive}} mode involves copying instructions and responses verbatim to ask questions and complete tasks, or waiting for the next guidance. The \textbf{\textit{active}} mode requires observable attention to part of knowledge, such as asking specific questions and critically analyzing responses. The \textbf{\textit{constructive}} mode further involves learner agency, such as strategically shaping the form of assistance and independently creating new artifacts from AI responses. We exclude the interactive mode from the original ICAP framework, as it does not map well to chatbot-based, task-oriented learning interactions~\cite{weatherholtz2025cognitive}. By combining the three engagement modes in help-seeking and response-use, we derived nine distinct reliance patterns (Figure~\ref{fig:teaser}).

\subsubsection{Knowledge Contexts}
Knowledge contexts correspond to the knowledge components addressed in each segment. We characterize these contexts along two dimensions. \textbf{Knowledge mastery} indicates whether a student has \textbf{\textit{acquired}} or \textbf{\textit{undeveloped}} a particular knowledge component. \textbf{Knowledge significance} distinguishes between \textbf{\textit{focal}} knowledge, which is essential for achieving the learning objective, and \textbf{\textit{supporting}} knowledge, which is secondary yet necessary for completing the learning activity. 

For example, in a statistics learning activity that uses Python to teach hypothesis testing, focal knowledge components may include understanding the meaning of a p-value and evaluating statistical hypotheses. A student may already have acquired or undeveloped these components prior to the activity. In contrast, Python programming is considered supporting knowledge, as it facilitates learning the focal concepts of the t-test but is not directly relevant to them.

To capture these knowledge contexts, researchers should administer a pre-test (e.g., one multiple-choice question per knowledge component) to assess students' prior knowledge and clearly specify the learning objectives to students in advance.

\subsection{Classifying Reliance Patterns}
\begin{figure}[ht]
\includegraphics[width=\linewidth]{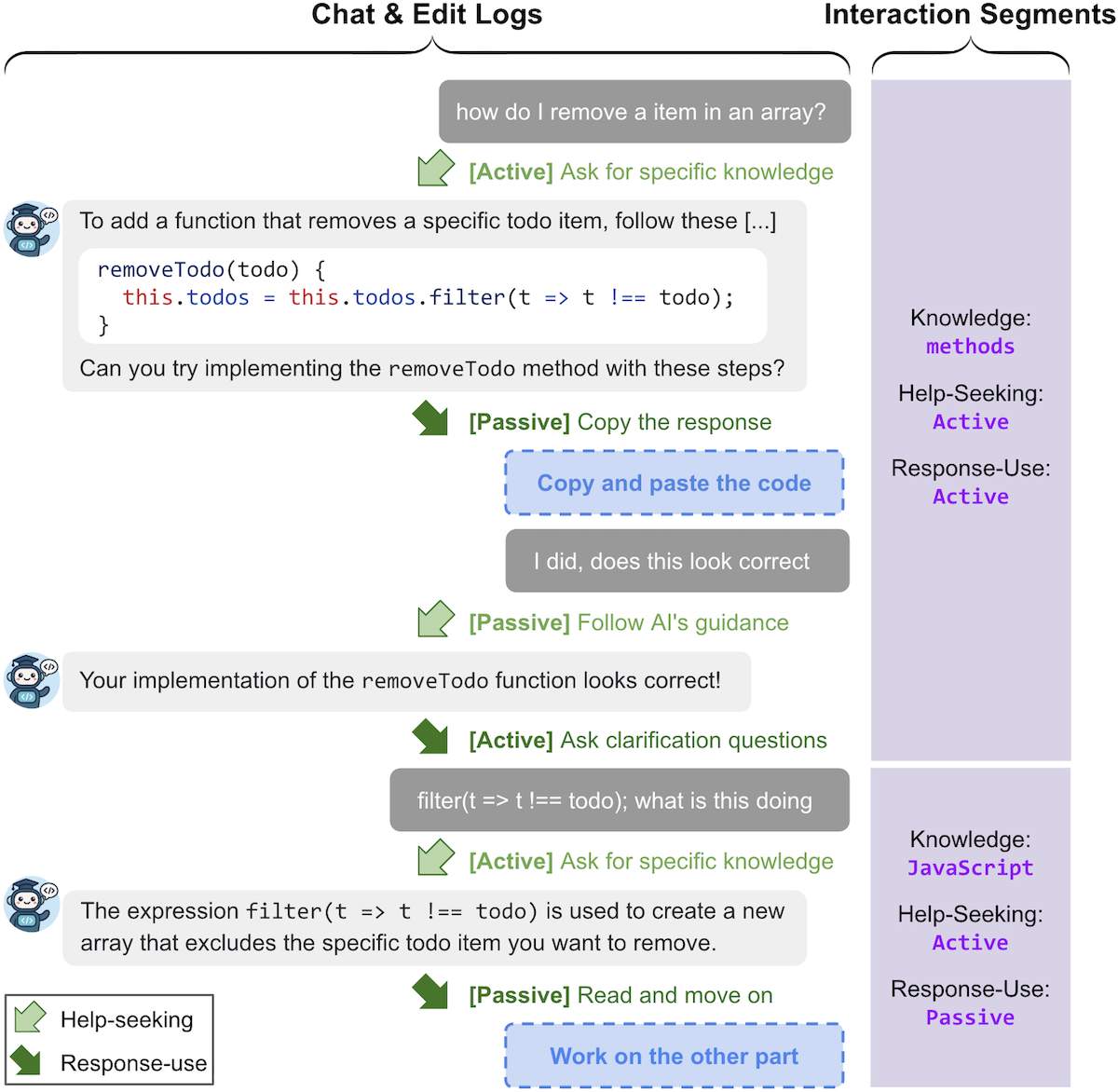}
    \caption{An illustration of how to classify reliance patterns in a programming learning activity. Purple boxes indicate interaction segments that group chat messages and code edit logs. Green annotations show engagement modes at the individual message level, which are aggregated into a single reliance pattern for each interaction segment.}
    \label{fig:example_conversation}
    \Description{A diagram showing how chat and edit logs are segmented into interaction segments using the framework. The left side displays a sequence of student-chatbot chat and edit logs, with each student message and edit action (e.g., copy and paste the code) annotated by Help-Seeking and Response-Use engagement levels (Passive, Active, or Constructive). A dashed line separates two interaction segments. The right side summarizes each segment with its associated knowledge context, Help-Seeking mode, and Response-Use mode. In the first segment, the knowledge context is methods, with Active Help-Seeking and Active Response-Use. In the second segment, the knowledge context is JavaScript, with Active Help-Seeking and Passive Response-Use.}
\end{figure}

We provide guidelines for classifying reliance patterns in practice. First, an interaction segment may exhibit multiple indicators of different reliance patterns. In such cases, classifiers (human or automated) should prioritize annotating the less-reliant modes (i.e., constructive) that are assumed to be rarer, thereby capturing more nuanced forms of reliance. For example, in Figure~\ref{fig:example_conversation}, the first interaction segment is annotated as \textit{Active} because it contains evidence of active engagement, even though other parts of the segment display \textit{Passive} behaviors.

Second, classifiers should consider adjacent interaction segments when determining the engagement mode. Engagement with a given knowledge component may be revealed in its subsequent segment rather than within itself. For instance, in Figure~\ref{fig:example_conversation}, the student initially discusses one knowledge component (``methods'') and then, in the subsequent segment, extends the conversation to a related component (``JavaScript''). Because the student builds on the initial explanation by asking follow-up questions, the first interaction segment is annotated as \textit{Active} response-use. In contrast, the second is annotated as \textit{Passive} response-use, as the student neither asked further questions nor modified the relevant part of the artifact.

%% file: table/codebook_help_seeking.tex
\begin{table*}[htp]
\caption{The codebook for classifying engagement modes in help-seeking. The examples show only a few cases, and users of the framework should adapt the interpretation of the modes to subject-specific interactions.}
\Description{A table with three columns: Help-Seeking Modes, Examples, and Subject-Specific Examples. The rows are grouped into three engagement levels, each with example behaviors. Passive includes seeking generic help, following AI's guidance, and requesting direct answers. Active includes seeking help for specific issues and asking specific knowledge questions. Constructive includes confirming a hypothesis, asking for a hint, and asking for an example. Subject-specific examples span programming, writing, and science domains, each illustrating student-chatbot interaction sequences.}
\label{table:codebook_help_seeking}
\begin{tabular}{lll}
\hline
\multicolumn{1}{l}{\textbf{Help-Seeking Modes}} &
  \textbf{Examples} &
  \textbf{Subject-specific Examples} \\ \hline
\multirow{3}{*}{\begin{tabular}[c]{@{}l@{}}Passive:\\ A learner signals\\ when help is needed\end{tabular}} &
  Seek generic help &
  \begin{tabular}[c]{@{}l@{}}{[}Writing{]} \raisebox{-0.25em}{\twemoji[scale=0.4]{student}} Help me write an essay.\\ {[}Programming{]} \raisebox{-0.25em}{\twemoji[scale=0.4]{student}} Complete function foo(). (copied from instruction)\end{tabular} \\ \cline{2-3} 
 &
  Follow AI's guidance &
  \begin{tabular}[c]{@{}l@{}}{[}Programming{]} \raisebox{-0.25em}{\twemoji[scale=0.4]{robot}}  Do you want me to write code for step 3? → \raisebox{-0.25em}{\twemoji[scale=0.4]{student}} Yes.\\ {[}Writing{]} \raisebox{-0.25em}{\twemoji[scale=0.4]{student}} I fixed the grammar error as you said.\end{tabular} \\ \cline{2-3} 
 &
  Request direct answers &
  {[}Programming{]} \raisebox{-0.25em}{\twemoji[scale=0.4]{student}} Give me code for adding an element. \\ 
\hline
\multirow{2}{*}{\begin{tabular}[c]{@{}l@{}}Active:\\ A learner specifies\\ what help is needed\end{tabular}} &
  \begin{tabular}[c]{@{}l@{}}Seek help for specific \\ issues\end{tabular} &
  \begin{tabular}[c]{@{}l@{}}{[}Programming{]} \raisebox{-0.25em}{\twemoji[scale=0.4]{student}} How do I resolve unexpected identifier error when trying to \\ pass id as a parameter?\\ {[}Science{]} \raisebox{-0.25em}{\twemoji[scale=0.4]{student}} My experiment result does not support what the textbook says.\end{tabular} \\ \cline{2-3} 
 &
  Ask specific knowledge &
  \begin{tabular}[c]{@{}l@{}}{[}Writing{]} \raisebox{-0.25em}{\twemoji[scale=0.4]{student}} What are rhetorical questions?\\ {[}Programming{]} \raisebox{-0.25em}{\twemoji[scale=0.4]{student}} What does filter() do?\end{tabular} \\ 
\hline
\multirow{3}{*}{\begin{tabular}[c]{@{}l@{}}Constructive:\\ A learner strategizes \\ how help is delivered\end{tabular}} &
  Confirm a hypothesis &
  {[}Science{]} \raisebox{-0.25em}{\twemoji[scale=0.4]{student}} Sublimation is from solid to air, right? \\ \cline{2-3} 
 &
  Ask for a hint &
  \begin{tabular}[c]{@{}l@{}}{[}Programming{]} \raisebox{-0.25em}{\twemoji[scale=0.4]{student}} Could you give me a skeleton to start from?\\ {[}Science{]} \raisebox{-0.25em}{\twemoji[scale=0.4]{student}} Could you give me insight on how to approach this problem?\end{tabular} \\ \cline{2-3} 
 &
  Ask for an example &
  {[}Programming{]} \raisebox{-0.25em}{\twemoji[scale=0.4]{student}} Do not write the entire code. Show me examples. \\ \hline
\end{tabular}
\end{table*}

%% file: table/codebook_response_use.tex
\begin{table*}[htp]
\caption{The codebook for classifying engagement modes in response-use. To classify engagement modes, researchers should ideally observe how learners edit artifacts.}
\label{table:codebook_response_use}
\begin{tabular}{lll}
\hline
\textbf{Response-Use Modes} &
  \textbf{Examples} &
  \textbf{Subject-specific Examples} \\ \hline
\multirow{3}{*}{\begin{tabular}[c]{@{}l@{}}Passive:\\ A learner accepts \\ AI's responses as given\end{tabular}} &
  \begin{tabular}[c]{@{}l@{}}Copy the response\end{tabular} &
  {[}Programming{]} \raisebox{-0.25em}{\twemoji[scale=0.4]{student}} \{paste the code given by AI into a code editor\} \\ \cline{2-3} 
 &
  Write over the response &
  \begin{tabular}[c]{@{}l@{}}{[}Programming{]} \raisebox{-0.25em}{\twemoji[scale=0.4]{robot}} \{provide answer code\} → \raisebox{-0.25em}{\twemoji[scale=0.4]{student}} \{type each character of \\ the code manually\}\end{tabular} \\ \cline{2-3} 
 &
  Read and move on &
  \begin{tabular}[c]{@{}l@{}}{[}Writing{]} \raisebox{-0.25em}{\twemoji[scale=0.4]{robot}} A rhetorical question creates a dramatic effect → \raisebox{-0.25em}{\twemoji[scale=0.4]{student}} \{read \\ the response and do other tasks\}\end{tabular} \\ \hline
\multirow{2}{*}{\begin{tabular}[c]{@{}l@{}}Active:\\ A learner analyzes \\ AI's responses\end{tabular}} &
  Correct artifact &
  {[}Writing{]} \raisebox{-0.25em}{\twemoji[scale=0.4]{robot}} \{point out grammar error\} → \raisebox{-0.25em}{\twemoji[scale=0.4]{student}} \{fix the error in a draft\} \\ \cline{2-3} 
 &
  Customize the response &
  \begin{tabular}[c]{@{}l@{}}{[}Programming{]} \raisebox{-0.25em}{\twemoji[scale=0.4]{robot}} \{provide answer code\} → \raisebox{-0.25em}{\twemoji[scale=0.4]{student}} \{make changes to variable \\ names, dummy data, etc.\}\end{tabular} \\ \cline{2-3} 
 &
  Ask clarification questions &
  \begin{tabular}[c]{@{}l@{}}{[}Science{]} \raisebox{-0.25em}{\twemoji[scale=0.4]{robot}} The second law: F=ma → \raisebox{-0.25em}{\twemoji[scale=0.4]{student}} What does each symbol \\ represent?\end{tabular} \\ \hline
\multirow{3}{*}{\begin{tabular}[c]{@{}l@{}}Constructive:\\ A learner adapts \\ AI's responses\end{tabular}} &
  Experiment with artifacts &
  \begin{tabular}[c]{@{}l@{}}{[}Programming{]} \raisebox{-0.25em}{\twemoji[scale=0.4]{robot}} \{provide answer code\} → \raisebox{-0.25em}{\twemoji[scale=0.4]{student}} \{change the code to test \\ hypothesis or new knowledge\}\end{tabular} \\ \cline{2-3} 
 &
  \begin{tabular}[c]{@{}l@{}}Create artifacts by oneself\end{tabular} &
  \begin{tabular}[c]{@{}l@{}}{[}Science{]} \raisebox{-0.25em}{\twemoji[scale=0.4]{robot}} \{provide subgoals to solve a problem\} → \raisebox{-0.25em}{\twemoji[scale=0.4]{student}} \{implement each \\ subgoal\}\end{tabular} \\ \cline{2-3} 
 &
  Ask constructive questions &
  \begin{tabular}[c]{@{}l@{}}{[}Science{]} \raisebox{-0.25em}{\twemoji[scale=0.4]{robot}} Matters can have three states → \raisebox{-0.25em}{\twemoji[scale=0.4]{student}} What is the state of jelly? \\ Solid or liquid?\end{tabular} \\ \hline
\end{tabular}
\end{table*}

%% file: section/40_study.tex
\section{Proof-of-Concept Study}
We demonstrate the use of \framework{}. This illustrative application can validate its efficacy in distinguishing nuanced reliance patterns and its scalability for such analysis using nonintrusively collected data. For example, if we observe different reliance patterns across students or knowledge contexts, we take it as evidence that \framework{} is useful for distinguishing nuanced reliance. We explore the following research questions.

\begin{enumerate}[label=\textbf{RQ\arabic*.}]
    \item How does \framework{} reveal different reliance patterns across students and knowledge contexts?
    \item What challenges do students face in managing their reliance during help-seeking and response-use?
    \item How does \framework{} enable reliance pattern analysis at scale?
\end{enumerate}

We conducted a data-collection study in a college-level web programming course. Students in the course completed a programming activity using an assistive chatbot to learn Vue.js. We collected and classified students' chat logs and code edit logs. To address RQ1, we conducted statistical analyses; for RQ2, we surveyed students; for RQ3, we benchmarked automatic classification.

\subsection{Participants}
The study involved 91 undergraduate students majoring in computer science, electrical engineering, and related fields. The students were taking an introductory web programming course in which they learned HTML, JavaScript, and Vue.js. A week before the study (academic week 11), students learned Vue.js in class. The students participated in the study as part of an assignment that affected their final grades. We collected students' chat and code edit logs with informed consent and IRB approval. We excluded two students who declined to share data from the analysis.

\input{table/study_procedure}

\begin{figure*}[ht]
\includegraphics[width=\textwidth]{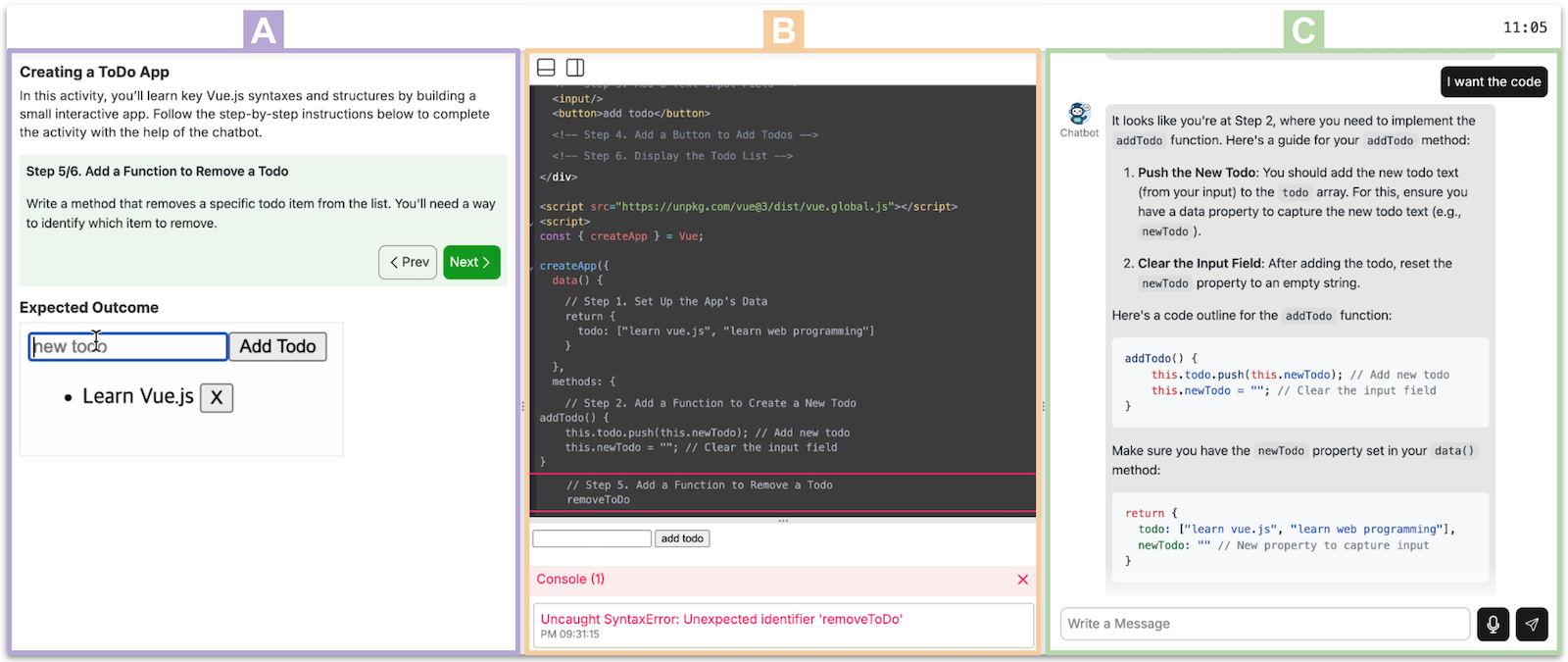}
    \caption{The interface used in the data collection study. The activity's learning objective and step-by-step guidance are always displayed on the left. Students could use a code editor to work and check results and console errors in real-time in the middle section. Students could also seek help from a chatbot on the right. A timer is located in the top-right corner.}
    \label{fig:interface}
    \Description{A system interface divided into three panes labeled A, B, and C. Pane A on the left displays the activity instructions, showing the title "Creating a ToDo App," a description of the learning objective, the current step, and an expected outcome. Pane B in the middle contains a code editor, a live preview area at the bottom with an input field and an add todo button, and a console section displaying a syntax error. Pane C on the right shows a conversation with a chatbot where the student asks for code, and the chatbot responds with step-by-step guidance and code snippets.}
\end{figure*}

\subsection{Procedure}
We conducted the study in an asynchronous, remote setting. The students visited our website on one of the nine days and spent an average of 30 minutes (SD = 15) without interruption. The study consisted of six stages, as shown in Table~\ref{table:study_procedure}. The materials used in the study are available in the supplementary materials\footnote{\url{https://osf.io/27ec5/overview?view_only=a8234a17f908464297d35d5ca1ef476c}}.

Before students interacted with a chatbot, we measured their prior knowledge (Stage 1 in Table~\ref{table:study_procedure}). The pre-test consisted of 10 multiple-choice questions, each with a corresponding Vue.js knowledge component and an ``I don't know'' option to minimize random guessing. The students were informed that their pre-test score would not affect their course grade. The pre-test had a 5-minute time limit, and the website displayed an alert prompting submission. 
The students then reported their self-efficacy, intrinsic motivation, extrinsic motivation, and metacognition (Stage 2) by answering 12 7-point Likert-scale questions~\cite{broadbent2023self}.

The students then spent 20 minutes on a learning activity (Stages 3 \& 4). The activity involved creating a To-Do app that taught them how to set up a Vue.js app, display data reactively, and handle user input events. The students used an interface shown in Figure~\ref{fig:interface}. On the left (A), students were instructed on the learning objective and step-by-step guidance to create the app with a demo video of the outcome. We decided to provide the six-step guidance to prevent students from requesting AI to generate the entire code at once. By having students work on their code incrementally, we aimed to observe multiple rounds of their self-regulation during their learning. Students could move between the steps as long as their code did not produce any console errors. In the middle (B), students were given a skeleton code and could edit it and check the outcome. For each step, the code editor highlighted the relevant part of the code. On the right (C), students could chat with an AI chatbot, starting with the message, ``Feel free to ask for any help!'' The chatbot could access the list of step-by-step guidance, answer code, students' current step, and their code. The chatbot was powered by gpt-4o-mini-2024-07-18 at the temperature set to 0 and had a minimal safeguard prompt to elicit students' natural usage of AI; however, we had to add an instruction that makes the chatbot refrain from giving answer code at once to avoid cases where students ask for a hint, but the chatbot gives direct answers.

After the learning activity, students took a post-test consisting of the same multiple-choice questions as the pre-test (Stage 5). The post-test also had a 5-minute time limit, and students were informed that their test score would impact their course grade. To further prevent random guessing, we informed students that we would deduct 1 point for incorrect answers. At the end (Stage 6), students answered four free-form questions about their challenges in forming appropriate reliance on AI.

\subsection{Data Logging}
For our analysis, we collected timestamped chat messages between students and the chatbot, along with snapshots of students' code at each message. We also logged students' code edits with timestamps by capturing code snapshots after 3 seconds following any new modification. To identify copy-and-paste, the system recorded additional code-edit logs whenever a single edit exceeded two characters. In addition, we logged the time students spent at each stage.

\subsection{Measures}
We measured the following to answer the research questions. 

\subsubsection{Knowledge components and reliance patterns.} 
Three authors collaboratively classified the knowledge components and reliance patterns associated with the messages and code edit logs. The authors conducted four rounds of classification and resolved disagreements after each round. Inter-rater reliability in the fourth round was acceptable~\cite{demszky2021measuring}, with agreement rates of 80.7\% for knowledge components, 83.6\% for help-seeking, and 71.6\% for response-use.

\subsubsection{Pre- and post-test.} Students received 1 point for each correct answer. There were 10 multiple-choice questions, and students could earn up to 10 points. Contrary to what was communicated to the students, we did not deduct points for incorrect answers.

\subsubsection{Self-regulation.} We summed the 7-point Likert ratings to create a 21-point scale for each of the four learning attributes. Questionnaire items are phrased positively so that high ratings correlate with strong self-regulation. The internal consistency (i.e., Cronbach's alpha) of the responses is high (self-efficacy: .94, intrinsic motivation: .91, extrinsic motivation: .73, metacognition: .80).

\subsubsection{Survey.} Students answered four open-ended questions about their understanding of the activity's learning objective, strategies for using the chatbot, and challenges in managing reliance during their help-seeking and response-use. We used affinity diagramming to identify students' challenges from their responses.

%% file: table/study_procedure.tex
\begin{table}[htp]
\caption{The procedure of the study. The numbers in parentheses indicate the time, in minutes, assigned to each stage.}
\Description{A table with two columns: Stage and Activity. Stage 1 (5 minutes) is a pre-test with 10 multiple-choice questions. Stage 2 is a self-regulation questionnaire. Stage 3 is an interface tutorial. Stage 4 (20 minutes) is learning Vue.js with a chatbot. Stage 5 (5 minutes) is a post-test identical to the pre-test. Stage 6 is a post-task survey. Stages 2, 3, and 6 do not have specified time limits.}
\label{table:study_procedure}
\begin{tabular}{lll}
\cline{1-2}
\textbf{Stage} & \textbf{Activity}                       &  \\ \cline{1-2}
1 (5)          & Pre-test (10 multiple-choice questions) &  \\
2              & Self-regulation questionnaire           &  \\
3              & Interface tutorial                      &  \\
4 (20)         & Learning Vue.js with a chatbot          &  \\
5 (5)          & Post-test (identical to pre-test)       &  \\
6              & Post-task survey                        &  \\ \cline{1-2}
\end{tabular}
\end{table}

%% file: section/50_results.tex
\section{Findings}
Our results provide evidence supporting \framework{}'s efficacy in (1) distinguishing nuanced reliance patterns between focal and supporting knowledge, (2) providing a holistic view of students' challenges, and (3) enabling scalable reliance analysis.

We collected 79 chat logs comprising 1,362 messages, 2,708 code-edit logs, and 427 interaction segments, annotated with associated knowledge components and engagement modes during help-seeking and response-use. This dataset excludes 11 chat logs in which students copied code from an external source---presumably a general-purpose chatbot---rather than using our chatbot. The average pre-test and post-test scores were 6 ($SD = 3$) and 9 ($SD = 1$), respectively, with students showing a statistically significant improvement on the post-test (paired t-test, $p < .001$). Students' average self-reported self-regulation scores, each on a 21-point scale, were as follows: self-efficacy ($M=16, SD=3.6$), intrinsic motivation ($M=14, SD=4.2$), extrinsic motivation ($M=12, SD=4.5$), and metacognition ($M=17, SD=3.6$). We report analyses addressing each research question. We use the notation \{\textit{help-seeking mode}\}\_\{\textit{response-use mode}\} throughout this section to concisely refer to each of the nine reliance patterns.

\subsection{RQ1: Students' Reliance Patterns}
\begin{figure}[htp]
\includegraphics[width=\columnwidth]{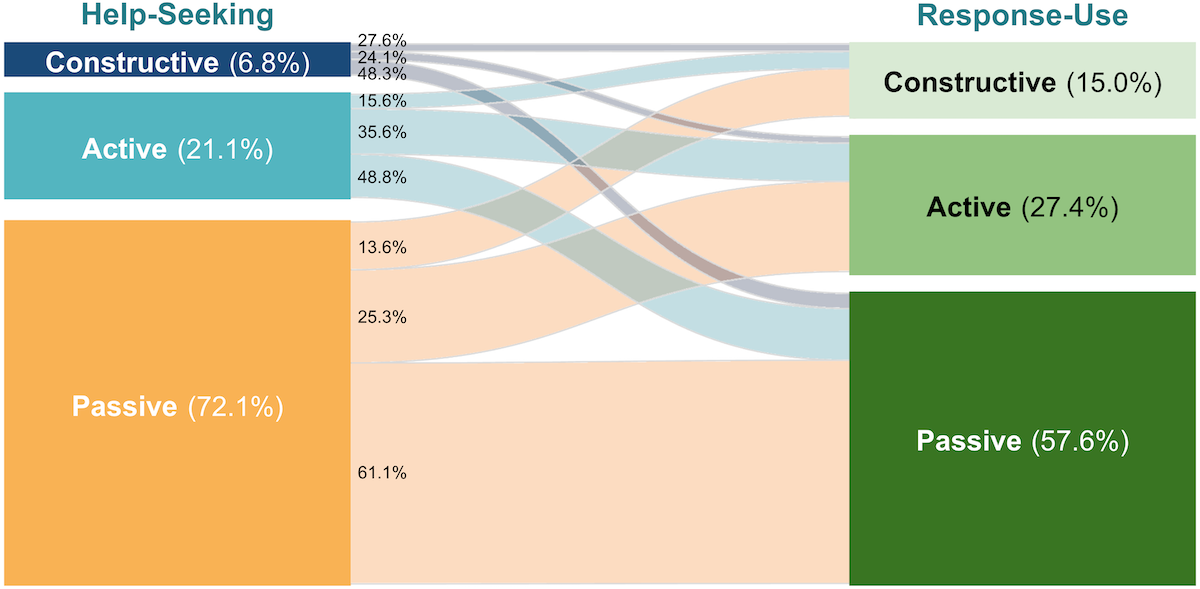}
    \caption{Flow of engagement modes from help-seeking to response-use, with band widths representing proportions of interaction segments.}
    \label{fig:distribution}
    \Description{A Sankey diagram showing the flow of engagement levels from Help-Seeking on the left to Response-Use on the right. Passive Help-Seeking is the most prevalent at 72.1\%, followed by Active at 21.1\% and Constructive at 6.8\%. For Response-Use, Passive is the most common at 57.6\%, followed by Active at 27.4\% and Constructive at 15.0\%. The largest flow is from Passive Help-Seeking to Passive Response-Use at 61.1\%. Active Help-Seeking most frequently flows to Active Response-Use at 48.8\%, and Constructive Help-Seeking most frequently flows to Constructive Response-Use at 48.3\%.}
\end{figure}

We first present the overall distribution of reliance patterns (Figure~\ref{fig:distribution}). Among the nine patterns, \textit{Passive\_Passive} was the most prevalent, accounting for 44.0\% of the 427 interaction segments. A common instance of this pattern involved students copying activity instructions into the chatbot as queries and then directly pasting the chatbot's responses back into their code editors (46 segments). Of the 427 interaction segments, 81\% involved \textit{Focal} knowledge, and 54\% of these segments involved \textit{Undeveloped Focal} knowledge. This distribution indicates that students relied on the chatbot to reproduce knowledge they already acquired as frequently as they used it to learn new knowledge.

\textbf{Active response-use correlates with active help-seeking.}
We conducted a Somers' D test to examine the relationship between engagement modes in help-seeking and response-use. The results revealed a significant positive association between the two actions ($D = .092, p = .024$), indicating that more active modes of help-seeking tend to co-occur with more active response-use behaviors. One possible explanation was that this relationship reflects stable learner traits (e.g., more motivated students engage more deeply across interactions). However, engagement modes in either help-seeking or response-use were not significantly associated (determined by Somers' D test) with any of the self-reported measures---self-efficacy, intrinsic motivation, extrinsic motivation, or metacognition, suggesting that the observed pattern cannot be fully explained by individual differences captured in our surveys.

As an alternative, we speculate that more specific questions elicit finer-grained responses that promote deeper engagement during integration. For instance, one student (S10) asked focused questions about isolated concepts (e.g., pushing an array, creating an object, defining a method) and chose to synthesize them into their code incrementally rather than requesting a combined answer. In contrast, students who posed broad or underspecified questions often received complete solution code, which they presumably struggled to analyze and adapt to their existing code, and were more likely to copy directly or even overwrite.

\begin{figure}[htp]
\includegraphics[width=\columnwidth]{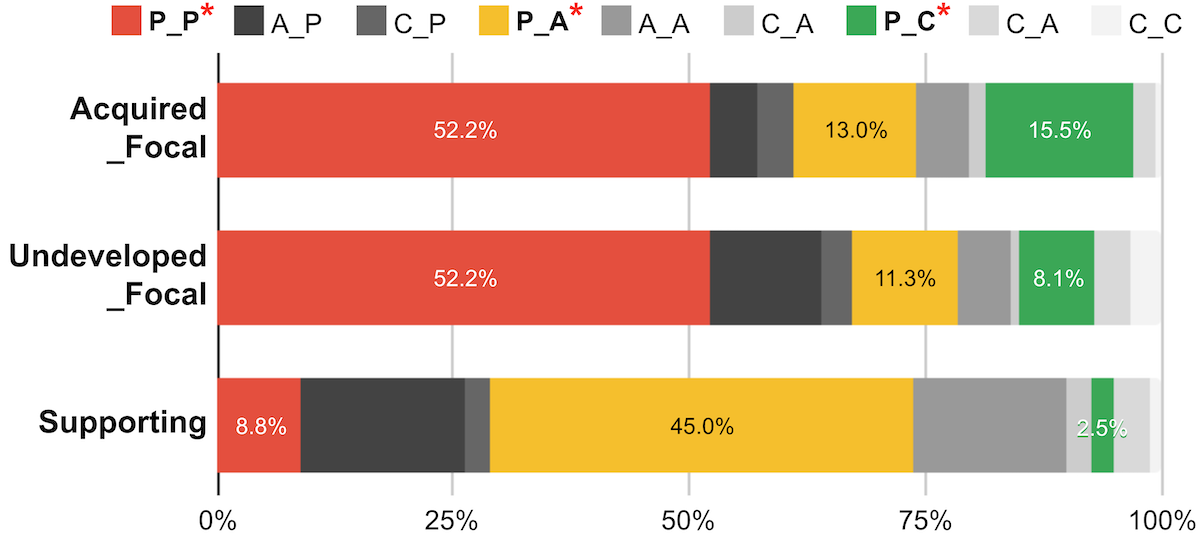}
    \caption{Distribution of reliance patterns across three knowledge contexts. Engagement modes are labeled by their first letter (e.g., P for \textit{Passive}). The proportions of P\_P, P\_A, and P\_C differ significantly between \textit{Focal} and \textit{Supporting}.}
    \label{fig:distribution_knowledge_contexts}
    \Description{A stacked horizontal bar chart with three rows representing knowledge contexts: Acquired Focal, Undeveloped Focal, and Supporting. Each bar shows the proportion of nine reliance patterns, labeled as combinations of Help-Seeking and Response-Use levels (e.g., P_P for Passive Help-Seeking and Passive Response-Use). Acquired Focal and Undeveloped Focal show similar distributions, both dominated by P_P at 52.2\%, with notable portions of P_A (13.0\% and 11.3\%) and P_C (15.5\% and 8.1\%). Supporting shows a distinct pattern, with P_A being the largest at 45.0\%, P_P reduced to 8.8\%, and P_C at 2.5\%.}
\end{figure}

\textbf{Knowledge mastery did not alter reliance behaviors.}
We conducted a MANOVA to examine whether the distribution of the nine reliance patterns differed across three knowledge contexts: \textit{Acquired\_Focal}, \textit{Undeveloped\_Focal}, and \textit{Supporting}. The independent variables were each student's proportional use of the nine patterns within each context. Because these proportions are compositional, we applied a centered log-ratio (CLR) transformation with a multiplicative replacement strategy~\cite{martin2003dealing} prior to analysis. The MANOVA revealed a significant effect of knowledge context (Pillai's trace: $F = 6.255, p < .001$). Follow-up ANOVAs with Games-Howell post-hoc tests showed no significant differences between \textit{Acquired\_Focal} and \textit{Undeveloped\_Focal}, whereas both differed from \textit{Supporting} in three reliance patterns (Figure~\ref{fig:distribution_knowledge_contexts}).

Similar reliance patterns across knowledge mastery are concerning because they suggest that students did not strategically allocate cognitive effort. Students could have relied more on the chatbot for \textit{Acquired} knowledge and devoted more effort to practicing \textit{Undeveloped} knowledge. This pattern may reflect students' limited awareness of knowledge gaps, insufficient metacognition to select effective strategies, or the chatbot's uniform interactions across mastery levels. 

\textbf{Active response-use may contribute more to learning than active help-seeking.}
We used an ordinary least squares (OLS) regression model to examine how different reliance patterns relate to students' post-test scores (Table~\ref{table:regression}). The model included students' pre-test scores and the total number of interaction segments as covariates, with the frequency of each reliance pattern within a chat as a predictor. We excluded 10 students who did not interact with the chatbot from this analysis. 

The coefficient for \textit{Passive\_Constructive} (0.371) was more positive than that for \textit{Active\_Passive} (-0.615), suggesting that deeper engagement during response-use may have a stronger association with learning than active help-seeking alone. However, because other coefficients did not reach statistical significance, we refrain from drawing definitive conclusions about the relative effectiveness of all reliance patterns.

\input{table/regression}

\begin{figure}[ht]
\includegraphics[width=\columnwidth]{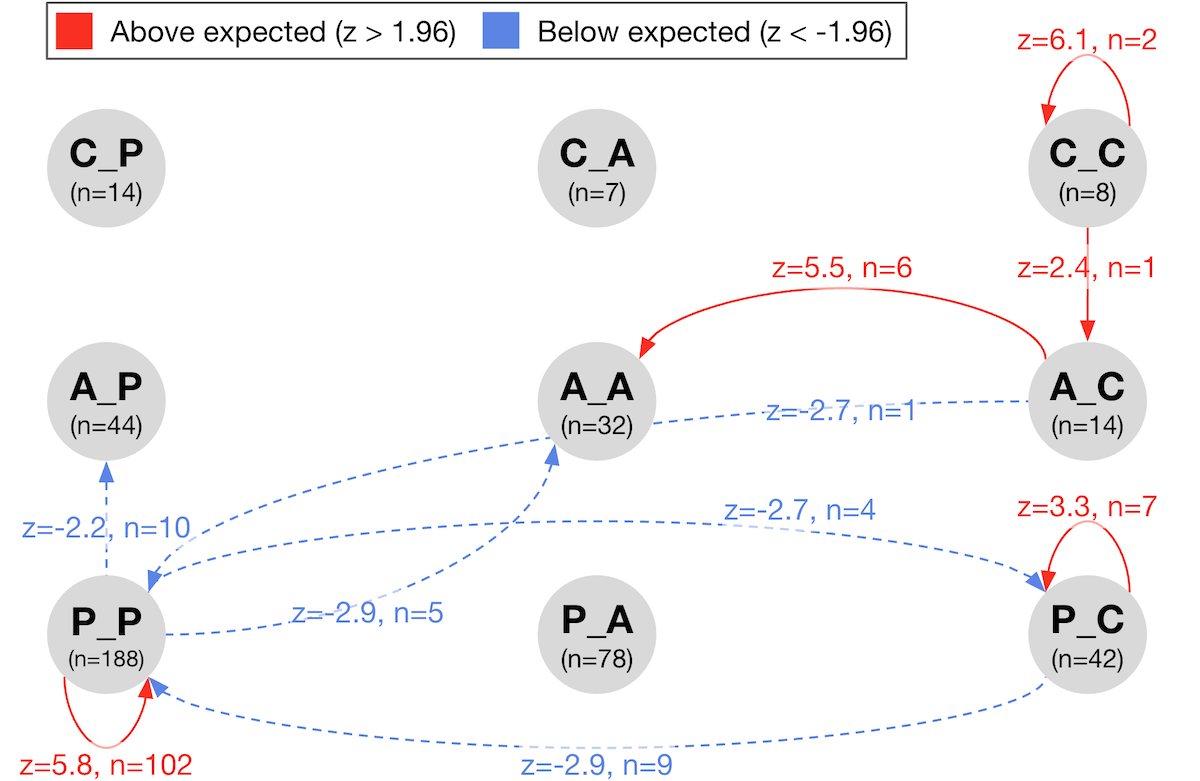}
    \caption{Red arrows show more common transitions between reliance patterns; blue arrows show less common ones.}
    \label{fig:lsa}
    \Description{A directed graph showing transitions between nine reliance patterns arranged in a three-by-three grid. Solid red arrows indicate statistically significant above-expected transitions (z-score greater than 1.96), and dashed blue arrows indicate below-expected transitions (z-score less than negative 1.96). Each arrow is labeled with its z-score and transition frequency. Notable above-expected transitions include self-loops at P_P and C_C, and a transition from A_C to A_A. Below-expected transitions include paths from P_P to higher engagement patterns such as A_P, A_A, and P_C, as well as from A_A and A_C to P_P.}
\end{figure}

\textbf{Students rarely shifted between reliance patterns.}
We conducted a lag sequential analysis (LSA) on reliance pattern sequences \cite{fan2026using, bakeman1997observing}, using adjusted residuals (z-scores) to identify transitions that occurred significantly more or less often than expected by chance ($z > 1.96$). The most frequent sequence was \textit{Passive\_Passive} $\rightarrow$ \textit{Passive\_Passive} ($z = 5.84, n = 102$) (Figure~\ref{fig:lsa}). 
Transitions from \textit{Passive\_Passive} to more engaged patterns, such as \textit{Active\_Active}, \textit{Active\_Constructive} or \textit{Active\_Passive}, were less frequent than expected. 
Another notable transition was \textit{Passive\_Constructive} $\rightarrow$ \textit{Passive\_Constructive} ($z = 3.27, n = 7$), suggesting that some students consistently relied on the chatbot for formulating help while writing their code independently. Together, these results indicate that reliance patterns tend to remain stable once established, with limited spontaneous shifts toward more engaged modes.

\subsection{RQ2: Challenges in Managing Reliance}
We analyzed students' survey responses to qualitatively examine their challenges in regulating help-seeking and response-use. Out of 91 students, only 2 reported not having a specific strategy. The majority of students were cautious about their reliance on AI and carefully limited its use to varying degrees. 

\textbf{Students struggled to pinpoint knowledge gaps and shape the chatbot's behavior for constructive help-seeking.}
17 of 91 students reported difficulty determining what to ask and how to ask. Some struggled to isolate the specific knowledge gaps within complex implementation steps that involved multiple knowledge components. As a result, they asked broad or underspecified questions, which led the chatbot to provide more hints or complete solutions than intended. Several students wanted to pose ``questions more geared toward learning rather than finding answers (S7),'' but faced ``wording challenges'' to prevent the chatbot from revealing answer code (S79). Twelve students struggled to align the chatbot with their personalized learning goals. Since students were using the chatbot for the first time, they were uncertain about its capabilities and how to ``write prompts to maximize its value (S86).'' The uncertainty limited their constructive help-seeking, such as steering the chatbot to provide brief answers to ``see what was wrong and what needed to be fixed and why (S58)'' or to provide more assistance with unfamiliar Vue syntax (S18).

\textbf{Artifact-centric or explanation-only responses made sense-making and adaptation difficult for constructive response-use.}
Eighteen students struggled to interpret the chatbot's responses. Several noted that the chatbot tended to provide the code ``with little explanation of the why or how (S66),'' which made it ``helpful for learning the syntax or structure, but less helpful for understanding the underlying reasoning (S41).'' A student (S58) preferred ``suggested edits rather than just given the answer.'' Five students reported difficulty applying abstract explanations to their own implementations. Students expressed a desire for ``more explanation of how [explanations] translates into wanted code (S55)'' Their comments highlight the need to leave room for engagement, but not so open-ended as to discourage it~\cite{chaiklin2003zone}. When the chatbot provided high-level guidance or solutions without decomposition, reconciling the response may have required overwhelming cognitive effort, thereby encouraging \textit{Passive} response-use, such as overwriting their code or requesting direct answers.

\subsection{RQ3: Scalability of \framework{}}
Automatic classification of reliance patterns can enable scalable analyses and support real-time, reliance-adaptive interventions for students~\cite{aleven2006toward, aleven2010automated}. One promising approach is to leverage large language models (LLMs) to automatically annotate reliance patterns. Accordingly, we evaluated how well lightweight and capable LLMs can classify reliance patterns, using our manually annotated dataset as a benchmark. The benchmark task was to classify engagement modes in help-seeking and response-use for each interaction segment. Each segment included the student's message, surrounding chatbot messages, the student's follow-up message, and all associated code edits and copy logs. We evaluated six models from three vendors (GPT, Gemini, and Qwen3) on 150 interaction segments, experimenting with multiple prompting strategies~\cite{brown2020language, wei2022chain}. Details of the prompts and evaluation settings are provided in the supplementary materials. We report results for the three models that are notable for their performance and size (Table~\ref{table:llm-performance}).

\input{table/llm-performance}

\textbf{LLMs can accurately detect \textit{Passive} modes. ($F1_{help-seeking} = .807$, $F1_{response-use} = .760$)}
For classifying help-seeking modes, Gemini-flash with 9 few-shot examples and a chain of thought achieved the best performance; for response-use, the zero-shot Gemini-pro was the best. Across models, LLMs tend to perform better at classifying help-seeking modes than response-use modes, likely because response-use classification requires interpreting code-edit logs in addition to conversational messages. Within each vendor, larger models generally yield higher performance. Notably, F1 scores are highest for detecting the \textit{Passive} mode in both help-seeking and response-use, and are comparable to human annotation agreement, which was 80\% and 70\%, respectively. This result is particularly promising for real-time reliance-adaptive support, as capturing passive behavior is critical for identifying over-reliance, which demands the greatest intervention.

\textbf{Few-shot examples improve LLM classification performance.}
Increasing the number of few-shot examples (from zero-shot to 3-shot and 9-shot) consistently improves classification performance across the five models, except Gemini-3-pro. This trend suggests that performance can be further enhanced as additional annotated data becomes available through our data collection system. Notably, reasonable performance is achieved with as few as nine shots, indicating that automatic reliance classification is both generalizable and accessible, without requiring large-scale data collection or activity-specific fine-tuning. Performance may be further improved by incorporating lightweight rule-based signals for certain patterns. For example, high textual similarity between a student's message and the activity instructions may indicate \textit{Passive} help-seeking, and the presence of copy logs immediately following a chatbot response may signal \textit{Passive} response-use.

%% file: table/regression.tex
\begin{table}[htp]
\caption{Ordinary least squares regression coefficients for the nine reliance patterns, predicting students' post-test score. Asterisks denote statistically significant coefficients ($p < .05$).}
\Description{A table with three columns: predictor variable, coefficient, and 95\% confidence interval. Eleven predictors are listed, including pre-test score, number of interaction segments, and nine reliance patterns. Two predictors show statistically significant coefficients: Active_Passive with a negative coefficient of negative 0.615, and Passive_Constructive with a positive coefficient of 0.371. The overall model has an R-squared of .341 and is statistically significant (p equals .004).}
\label{table:regression}
\begin{tabularx}{\linewidth}{lll}
\hline
                               & \textbf{Coefficient} & \textbf{95\% CI}     \\ \hline
Pre-test                       & 0.028               & {[}-0.085, 0.140{]}  \\
Number of Interaction segments & 0.023               & {[}-0.139, 0.185{]}  \\
Passive\_Passive                & -0.017              & {[}-0.207, 0.173{]}  \\
Active\_Passive                 & -0.615*             & {[}-0.980, -0.251{]} \\
Constructive\_Passive           & 0.397               & {[}-0.288, 1.083{]}  \\
Passive\_Active                 & 0.202               & {[}-0.184, 0.587{]}  \\
Active\_Active                  & -0.176              & {[}-0.568, 0.215{]}  \\
Constructive\_Active            & -0.243              & {[}-1.226, 0.741{]}  \\
Passive\_Constructive           & 0.371*              & {[}0.040, 0.701{]}   \\
Active\_Constructive            & -0.049              & {[}-0.640, 0.543{]}  \\
Constructive\_Constructive      & 0.153               & {[}-0.505, 0.811{]}  \\ \hline
\multicolumn{3}{r}{$R^2 = .341$, Adj. $R^2 = .228$, $p = .004$} \\
\hline
\end{tabularx}
\end{table}

%% file: table/llm-performance.tex
\begin{table}[htp]
\caption{Performance of LLMs in classifying reliance patterns from students' chat, code-edit, and copy logs. Bold values indicate the highest F1 score across models.}
\Description{A table comparing LLM classification performance across three models (gemini-3-pro-preview, gemini-3-flash-preview, and qwen3:8b) and four prompting strategies (Zero-Shot, 3-Shot, 9-Shot, and 9-Shot plus Chain-of-Thought). Performance is reported using F1 Passive and F1 Micro scores for both Help-Seeking and Response-Use classification. The highest Help-Seeking scores are achieved by gemini-3-flash-preview with 9-Shot plus Chain-of-Thought (F1 Passive of .807, F1 Micro of .720), while the highest Response-Use scores are achieved by gemini-3-pro-preview with Zero-Shot prompting (F1 Passive of .760, F1 Micro of .633).}
\label{table:llm-performance}
\begin{tabularx}{\linewidth}{llllll}
\hline
\multirow{2}{*}{\textbf{Model}} &
  \multirow{2}{*}{\textbf{Prompting}} &
  \multicolumn{2}{l}{\textbf{Help-Seeking}} &
  \multicolumn{2}{l}{\textbf{Response-Use}} \\
                           &              & $F1_{Passive}$   & $F1_{Micro}$     & $F1_{Passive}$   & $F1_{Micro}$    \\ \hline
\multirow{4}{*}{\begin{tabular}[c]{@{}l@{}}gemini-3\\ -pro\\ -preview\end{tabular}}   
                           & Zero-Shot    & .763          & .707          & \textbf{.760} & \textbf{.633} \\
                           & 3-Shot       & .723          & .673          & .722          & .580          \\
                           & 9-Shot       & .736          & .640          & .704          & .580          \\
                           & 9-Shot+CoT   & .702          & .627          & .678          & .580          \\ \hline
\multirow{4}{*}{\begin{tabular}[c]{@{}l@{}}gemini-3\\ -flash\\ -preview\end{tabular}}
                           & Zero-Shot    &.667           & .607          & .585          &.487 \\ 
                           & 3-Shot       & .700          & .667          & .593          & .473          \\
                           & 9-Shot       & .762          & .693          & .667          & .533          \\
                           & 9-Shot+CoT & \textbf{.807} & \textbf{.720} & .667          & .580          \\ \hline
\multirow{4}{*}{qwen3:8b}  & Zero-Shot    & .751          & .540          & .635          & .427          \\
                           & 3-Shot       & .733          & .540          & .628          & .427          \\
                           & 9-Shot       & .750          & .560          & .646          & .453          \\
                           & 9-Shot+CoT & .533          & .460          & .593          & .460          \\ \hline
\end{tabularx}
\end{table}

%% file: section/60_discussion.tex
\section{Discussion}

\subsection{Implications of \framework{}}
The significance of \framework{} lies in its interaction-centric design, which jointly models help-seeking and response-use to capture reliance holistically. Prior work has largely examined these behaviors in isolation---an approach suited to structured tutoring systems that constrain either how students ask for help or how they consume feedback, but ill-suited to the open-ended nature of generative AI chatbots. Our analysis shows that 46\% of interactions exhibit different engagement modes in help-seeking and response-use, including sharply contrasting patterns (e.g., \textit{Passive}\_\textit{Constructive}). For example, one student (S37) repeatedly engaged in passive help-seeking by submitting task instructions to the chatbot, yet constructively used its responses to independently write code. Such hybrid reliance patterns fall between full reliance and full independence and would be obscured by existing measures, despite being critical for understanding how reliance manifests in practice. While prior work suggests that constructive response-use may compensate for passive help-seeking~\cite{aleven2016help} and that guided learning can outperform self-exploration~\cite{yannier2020active}, \framework{} provides a systematic means to investigate the interplay between help-seeking and response-use.

The interaction model underlying \framework{} also provides both a lens for understanding students' challenges and a foundation for designing student-chatbot interactions. Existing conversational tutoring systems primarily focus on improving chatbot responses through step-by-step explanations and answer-avoidance guardrails. However, interactions with generative chatbots begin with open-ended help-seeking, which our findings show is closely associated with response-use. Designing pedagogical chatbots, therefore, requires not only better response generation but also organically integrated environmental support for self-regulated help-seeking~\cite{hooshyar2025problems}. Such support includes help-seeking training, question templates~\cite{king1994guiding}, and metacognitive tutors~\cite{aleven2006toward, alaimi2020pedagogical}.

We encourage future research to further extend and apply \framework{}. For example, the learning-context lens could be expanded to incorporate additional factors such as learners' need for cognition~\cite{buccinca2021trust} or interest in specific knowledge components. Examining longer sequences of reliance patterns over time may also yield insights into how reliance evolves during learning activities~\cite{amoozadeh2024student}. Future work could leverage automatic classifiers~\cite{weatherholtz2025cognitive} to operationalize \framework{} for real-time reliance tracking, enabling adaptive interventions that help address the long-standing assistance dilemma in AI-supported learning environments~\cite{mathan2018fostering, koedinger2007exploring, chen2025more}.

\subsection{Design Implications for Learning Systems}
We share several directions for designing future AI-assisted learning systems. Most college-level students in the survey reported awareness of the risks of over-reliance on AI and an intention to self-regulate their use of chatbots. However, students lacked adequate support for decomposing learning tasks and requesting assistance only for the specific components they had not yet mastered. These gaps indicate opportunities for intervention at both the learning-activity and chatbot-design levels.

At the level of learning activities, we note the value of incremental task design. In our study, the programming assignment was decomposed into six sub-activities, revealed sequentially. Notably, none of the 79 students asked the chatbot to generate the entire solution at once. This incremental structure likely helped students more accurately assess their knowledge state and created more opportunities to adjust their help-seeking and reliance on AI over time. Such task-level scaffolding may be more robust than improving chatbots only, as students can easily opt out of institutional chatbots in favor of general-purpose tools, whereas they cannot bypass incrementally revealed instructions.

At the level of chatbot design, systems can support self-regulation and effective question formulation by decomposing students' requests into subtasks and allowing learners to select which subtasks to receive help with. Such decomposition can elicit modular, digestible responses that encourage deeper engagement as students incrementally integrate them into their artifacts. Complementarily, metacognitive tutoring approaches that provide feedback on students' chatbot-interaction strategies may further promote productive AI use~\cite{roll2006help, aleven2006toward, jin2024teach}. Our survey results also indicate that students struggle to align chatbot behavior with their personal learning goals. To address this, chatbots could incorporate an explicit calibration stage---configured upfront or adjusted iteratively---to support goal-help alignment throughout the learning process.

%% file: section/70_limitation.tex
\section{Limitation and Future Work}
We note several limitations. First, students' reliance behaviors in our study may not fully generalize to interactions with commercial chatbots (e.g., ChatGPT), as our system employed customized prompts and students were aware that their interactions were logged. Although we designed appropriate incentives and clearly communicated learning objectives, time constraints and grading policies may have influenced students' reliance behaviors. Second, we do not establish strong statistical evidence linking reliance patterns to learning gains. Noise in self-reported measures and the use of a one-time knowledge assessment limited our ability to fit robust regression models. Establishing such relationships will require larger datasets and repeated measures. We encourage future work to apply \framework{} across varied and controlled learning settings to replicate and extend our findings and to more precisely characterize when and how reliance on AI supports learning.